# Nonlinear dynamics analysis of a membrane Stirling engine: Starting and stable operation

## Author

Fabien Formosa

## Affiliations

Laboratoire SYMME, Université de Savoie

BP 80439- 74944 ANNECY LE VIEUX CEDEX - FRANCE

Tel: +33 4 50 09 65 08

Fax: +33 4 50 09 65 43

Mail: fabien.formosa@univ-savoie.fr

## Abstract

This paper presents the work devoted to the study of the operation of a miniaturized membrane Stirling engine. Indeed, such an engine relies on the dynamic coupling of the motion of two membranes to achieve a prime mover Stirling thermodynamic cycle. The modelling of the system introduces the large vibration amplitudes of the membrane as well as the nonlinear dissipative effects associated to the fluid flow within the engine. The nonlinearities are expressed as polynomial functions with quadratic and cubic terms. This paper displays the stability analysis to predict the starting of the engine and the instability problem which leads to the steady state behaviour. The centre manifold - normal form theory is used to obtain the simplest expression for the limit cycle amplitudes. The approach allows the reduction of the number of equations of the original system in order to obtain a simplified system, without loosing the dynamics of the original system as well as the contributions of non-linear terms. The model intends to be used as a semi analytical design tool for the optimization of miniaturized Stirling machines from the starting to the steady operation.

**1. Introduction**

One of the new application of the Stirling engine is the free piston configuration proposed by W. Beale of the Ohio University by the end of the sixties [1]. Without any mechanical linkages between the piston and the displacer, this architecture allows the realization of fully sealed machine which leads to the most reliable machines compared to the more classical mechanical arrangements. In such a machine, the reciprocating motions of the piston and displacer are effected by the dynamical and the fluidic interactions. The operation of free piston Stirling engine (FPSE) results from a self starting characteristic followed by a stable operation stage when nonlinear phenomena modify the initial linear behaviour limited to the small amplitudes of the moving parts. In The nonlinearities are generally related to pressure drops when the working fluid goes back and forth within the heat exchangers and the regenerator and consequently allow the stable operation. Thus, the thermodynamic cycle can be achieved with appropriate parameters. The main advantages are: a simple mechanical design; minimal lateral forces on moving components which minimizes the engine wear; high energetic efficiency; great life time and low cost manufacture.

Without any mechanical linkage the thermodynamic optimization is a complex issue. Principal design parameters, swept and dead volume ratios, phase angle and speed of engine are coupled. In depth dynamical analysis is therefore necessary to account for design parameters effects on performances. Though a numerical model is to be considered usually, a semi analytical approach would allow an efficient preliminary design for a strongly coupled problem with numerous design parameters [2]. The design of such an engine requires to master complex thermal, mechanical and fluid dynamics interactions. In most analytical studies assumptions of the isothermal Schmidt cycle [3] are adopted then the traditional approach consists in linearizing of each force which acts on the moving parts and considering the whole problem as a coupled two degrees of freedom (DOF) system. One obtains therefore the dynamical equations in the neighbourhood of a stable operating point which leads to the dynamical and consequently the thermodynamic characteristics of the FPSE [4-6].

Few works deal with miniaturized Stirling machine for which the moving components are replaced by vibrating membranes. Bowman in [7] and [8] developed an analytical model to design membranes which match the dynamical criteria (notably swept volume and intrinsic dynamical characteristics). However, this analysis is restricted to linear behaviour whereas the geometric nonlinear phenomena appear very soon as a flexural behaviour is to be considered [9]. Under these conditions, the approach proposed by De Monte Refs. [4-5] and Rogdakis [10] which includes the nonlinear dissipation phenomena associated to the fluid flow is completed hereafter by the mechanical nonlinearity of each membrane. The later appears to be a crucial point to forecast the performances of a membrane Stirling engine.

During recent years, lots of works which aim at the understanding of the dynamic behaviour of systems with nonlinear phenomena have been developed. Perturbation methods, such as averaging or the methods of multiple scales have been used in many studies for simplification purpose of the effects of nonlinear phenomena which occur in many problems related to engineering and science [11]. The centre manifold approach is the one which leads to the most simplified model. It accounts for the local bifurcation behaviour in the neighbourhood of a fixed point of the nonlinear system [12]. The centre manifold approach reduces the number of equations of the original system in order to obtain a simplified system without loosing the dynamics of the original system as well as the effects of nonlinear terms [13]. Moreover, the normal form approach can be jointly used to eliminate as many nonlinear terms as possible through nonlinear successive changes of variable.

The present paper depicts the application of the centre manifold theory and the normal form approach to a membrane miniaturized Stirling engine (MMSE) under the Schmidt assumptions. The membrane micro engine arrangement and its expected behaviour are first introduced. Next, the dynamic of the membranes will be outlined. A first simplification method based on a modal projection leads to a set of two coupled second order differential equations which is representative of the system's dynamic. The coupling from the isothermal evolution assumption which accounts for the pressure to be analytically related to temperature and volume variations is established. The system stability is then assessed by the linearization of the equations at a fixed point and self starting conditions are outlined. Finally, the centre manifold and the normal form approaches will be used in order to predict limit cycle amplitudes and the estimation the performances of the engine eventually.

**2. Mechanical arrangement and qualitative working of a membrane miniature Stirling engine (MMSE)**

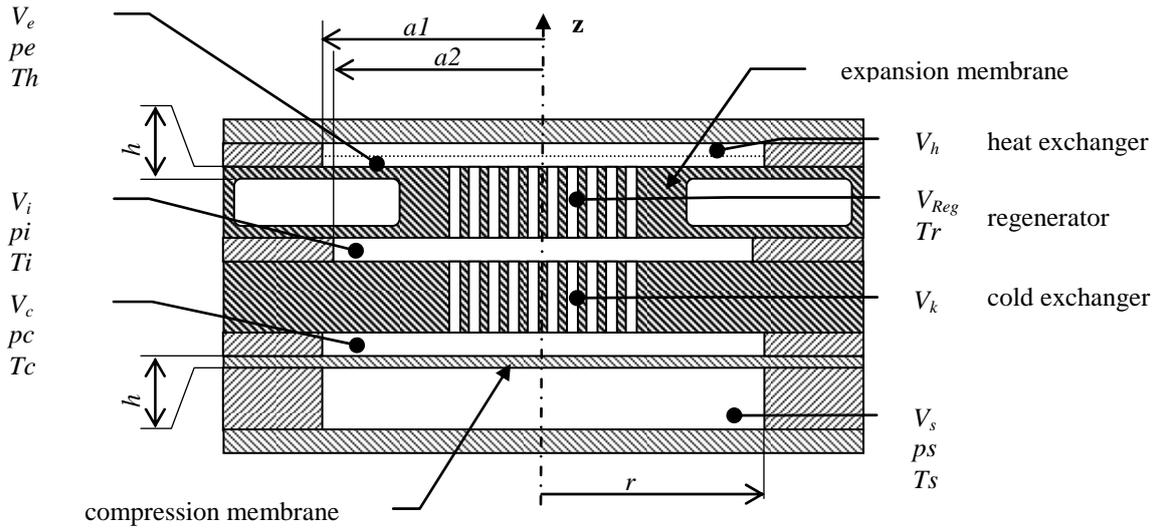

Fig. 1 Mechanical arrangement of the micro Stirling engine.

Based on the patent of a Stirling micro machine from Bowman [7], the Fig. 1 shows its generic mechanical arrangement as well as the relevant parameters for the study.

The operation of the engine is based on the bending of two membranes called $m_{exp}$ and $m_{comp}$ which can therefore modify the volumes of the chambers $V_e$, $V_i$ and $V_c$, $V_s$ respectively. The compression membrane denoted $m_{comp}$ ensures the role of the piston and the expansion membrane denoted $m_{exp}$ the role of the displacer of the Stirling machine. For an engine, the volume $V_e$ is at the highest temperature $Th$ whereas $V_c$ and $V_i$ are maintained to lower temperatures $Tc$ and $Ti$ thanks to the cold exchanger. On can note that we make a distinction between the spaces above and under the cold heat exchanger (with respect to the z axis shown on Fig. 1) in order to account for a pressure drop as well as a temperature difference between the two. A moving regenerator $V_{Reg}$ provides thermal insulation between $V_e$ and $V_i$ and allows better performances alternatively storing and releasing heat to the working fluid. The

Table **1** below gives the geometric characteristics used for this study.

Table 1

| Chambers volumes (cm³) | | | | | | Membranes radii (mm) | | | Height (µm) | |
| --- | --- | --- | --- | --- | --- | --- | --- | --- | --- | --- |
| | | | | | | $m_{comp}$ | $m_{exp}$ | | $m_{comp}$ | $m_{exp}$ |
| $V_h$ | $V_e$ | $V_{Reg}$ | $V_i$ | $V_k$ | $V_c$ | ra | a1 | a2 | hp | hd |
| 0.5 | 0.75 | 0.12 | 0.5 | 1.5 | 0.25 | 20 | 20 | 16 | 40 | 20 |

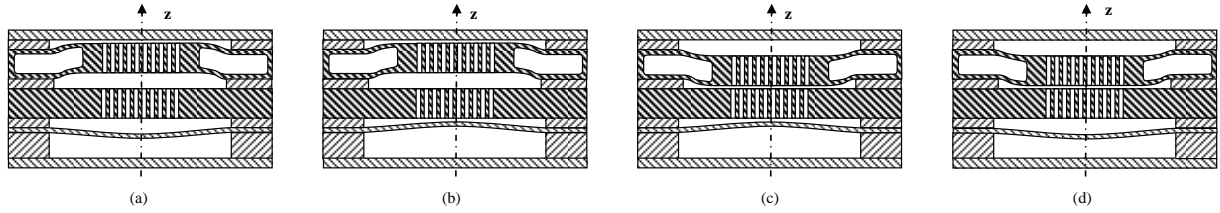

Fig. 2 Representative stages of operation.

The representative stages for the operation of the engine can be described from a reference position of the bending components. In the Fig. 2a, the compression membrane is at its lowest position as the expansion membrane is at its highest so both $V_c$ and $V_i$ are maximum and $V_e$ is minimal. In Fig. 2b, the upward motion of the compression membrane induces a global compression of the working fluid. The subsequent temperature increase is balanced by an outward heat flow at the cold exchanger. With a downward motion of the expansion membrane the working fluid is partly forced trough the regenerator. Therefore, heat energy is provided to the fluid by the regenerator material so it ideally emerges from it at the temperature $Th$ (Fig. 2c). The global pressure consequently rises as for a constant volume the temperature becomes higher. Submitted to this pressure, the compression membrane moves to its initial position and by doing this, some mechanical power is extracted from the engine. Spring forces lead to an out of phase motion of the expansion membrane (Fig. 2d). The heat source provides the amount of energy to ideally keep a constant high temperature $Th$ during the resulting global expansion. At the end of the expansion stage the expansion membrane is at its initial position and the regenerator material recovers the amount of heat energy harnessed by the fluid during the previous blow (from $V_i$ to $V_e$). In the ideal case, the fluid went back to the compression spaces ($V_c$ and $V_i$) at the temperature $Tc$.

## 3. Semi analytical model of the membranes

In order to simulate the global behaviour of the engine, one first interests in the modelling of the moving parts. For the considered micro engine the compression membrane is assumed to be a single thin, circular and clamped plate whereas the expansion membrane is supposed to be composed of two clamped circular shells linked by a rigid stem at their centre (see Fig. 1 and Fig. 13).

Haterbouch-Benamar [9], [14] developed a theoretical model based on Hamilton's principle and spectral analyse. A semi analytical relation between the eigenmodes, the eigenfrequencies and the amplitude of bending for large deflections can thus be established. This method has been successfully applied for numerous typologies of structures such as simply supported or clamped beams and plates [15-17]. The method is especially efficient not only to account for the amplitude-frequency relation but also represents the effect of large deflexion on the shape of the membrane. Because of the importance of the swept volume on the performance of the micro engine, this last result happens to be relevant in the design procedure. The main assumptions used for the modelling of the membrane are the following:

- Membranes are submitted to mechanical solicitations such as their strains are only axisymetric.
- The only displacement under consideration is the displacement in the z direction denoted w*(r).
- Plane stresses are considered (shear stresses are neglected).
- Rotational kinetic energy is neglected.

To the extent that the harmonic motion is supposed, strain and kinetic energy can be obtained using the basis of linear eigenmodes which form the spectral basis to express the transverse displacement. In the case when large deflexion is to be considered, nonlinear geometric behaviour leads to a tensor of the fourth order associated to in plane induced tension in addition to the classical stiffness and mass ones. It follows from the application of the Hamilton's principle the sets of algebraic nonlinear equations which can be numerically solved. Provided that the spectral basis can be written using Bessel's functions of the first and second kind, this approach has the advantage to provide a semi analytical model of the nonlinear dynamical behaviour for a membrane.

In the following, the method is applied to the compression and the expansion membranes although in this study, the restriction of the spectral basis to a single vector will be used in order to obtain only one degree of freedom (DOF) equation for each of the membrane such as:

$$\ddot{q}_1 m^*_{11} + q_1 k^*_{11} + 2 q_1^3 b^*_{1111} = p f^*_1 \qquad (1)$$

In which $\ddot{q}_1$ and $q_1$ are the acceleration and displacement response of the generalized coordinate and $p$ is a pressure applied to the membrane area. $m^*_{11}$, $k^*_{11}$, $b^*_{1111}$ and $f^*_1$ are the mass, linear stiffness, non-linear stiffness and reduced pressure coefficient terms which are evaluated from the spectral analysis of the membrane equilibrium equation (the method is described in Appendix).

It is worthy of note that the previous Eq. (1) can be seen as a conservative Duffing's oscillator from which the characteristic evolution of the resonance frequency of the membrane with respect to its deflexion is plotted in Fig. 3.

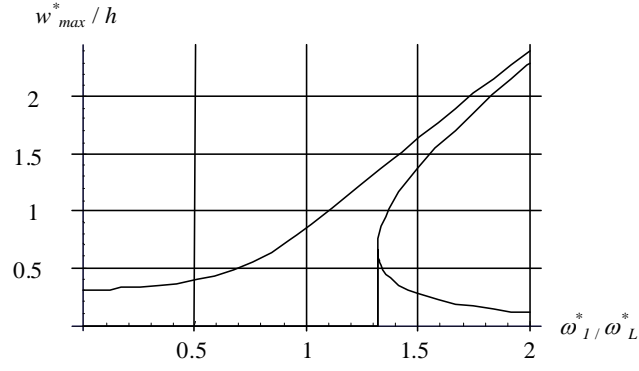

Fig. 3 Normalized resonance frequency with respect to normalized amplitude.

Yet, the single DOF model is relevant to represent the effect of the amplitude of deflexion on the first resonance frequency for a membrane under large strain assumption. Thorough study with larger spectral basis would not involve much theoretical difficulties but would consequently increase the number of DOF.

The development of the analytical method for the compression membrane and the expansion membrane is developed in Appendix.

## 4. Analytical model of the MMSE

The governing equations of the micro engine can thus be obtained using a two steps procedure. The first step as was shown before establishes a single DOF nonlinear equation for each of the membrane such that:

$$\ddot{q}_p + 2\,\xi_p \sqrt{\frac{k_p}{m_p}}\,\dot{q}_p + \frac{k_p}{m_p}\,q_p + 2\frac{b_p}{m_p}\,q_p^3 = (ps\text{-}pc)\frac{f_p}{m_p} \qquad (2)$$

$$\ddot{q}_d + 2\,\xi_d \sqrt{\frac{k_d}{m_d}}\,\dot{q}_d + \frac{k_d}{m_d}\,q_d + 2\frac{b_d}{m_d}\,q_d^3 = pi\frac{f2_d}{m_d} - pe\frac{f1_d}{m_d} \qquad (3)$$

With the following initial conditions:

$$q_p|_{t=0} = 0\,,\ \dot{q}_p|_{t=0} = 0 \ \text{and}\ q_d|_{t=0} = 0\,,\ \dot{q}_d|_{t=0} = 0 \qquad (4)$$

Where $(m_p, k_p, b_p)$ and $(m_d, k_d, b_d)$ are the mass, linear stiffness and non-linear stiffness terms which are evaluated from the spectral analysis of the compression membrane and the expansion membrane respectively. $pe$, $pi$, $pc$ and $ps$ are the gas pressure of the chambers as define in Fig. 1.

Eq. (2) stands for the dynamic behaviour of the compression membrane. $q_p$ is the displacement response of the compression membrane DOF and $\dot{q}_p$ is the derivative of $q_p$ with respect to time. $\xi_p$ is a modal damping coefficient added to represent the mechanical dissipative phenomena. It is worthy of note that the damping coefficient $\xi_p$ can represent the effects of a linear load applied to the micro engine. As an example, such a load can be induced by an electromagnetic transducer. As a consequence, its value will be chosen in a quite wide range. Set of Eqs. (2-4) can also be seen as two oscillators coupled by mechanical forces.

Eq. (3) stands for the dynamic behaviour of the expansion membrane. $\ddot{q}_d$, $\dot{q}_d$ and $q_d$ are the acceleration, the velocity and displacement response of the expansion membrane DOF, respectively. $\xi_d$ is a modal damping coefficient added to represent the mechanical dissipative phenomena.

For the second step, under usual assumptions of isothermal evolutions and ideal working fluid, the instantaneous pressures within the chambers can be analytically expressed. The total mass of gas inside the various chambers is defined as the sum of the mass of the gas contained in each of the chamber defined in Fig. 1:

$$M = m_e + m_h + m_{Reg} + m_i + m_k + m_c \tag{5}$$

If we substitute the expressions of the masses in Eq. (5) with the ideal gas law (e.g. $m_e = \frac{p\,V_e}{R\,Th}$), we obtain:

$$p = M\,R\,/\,(\frac{V_e}{Th} + \frac{V_h}{Th} + \frac{V_{Reg}}{Tr} + \frac{V_i}{Ti} + \frac{V_k}{Tc} + \frac{V_c}{Tc}) \tag{6}$$

where $R$ is the considered gas constant per unit of mass and $Tr$ is the mean temperature of the regenerator. As expected, the pressure depends on the motion of both membranes. As an example, the motion of the compression membrane modifies the pressure. Moreover, the different pressures of the chambers are linked by the pressure losses effects which occur in the heat exchangers as well as in the regenerator.

Thus denoting $p$ the mean global pressure, we get: $pe = p + \frac{1}{2}\,\Delta p_{Reg}$, $pc = p - \frac{1}{2}\,\Delta p_{Ech}$ and $pi = p - \frac{1}{2}\,\Delta p_{Reg} + \frac{1}{2}\,\Delta p_{Ech}$.

where $pe$, $pi$ and $pc$ are the pressures defined in Fig. 1.

The expressions of the pressure losses $\Delta p_{Reg}$ and $\Delta p_{Ech}$ are highly nonlinear relation in which the classical Reynolds number is the main variable. To simplify the approach, cubical nonlinearities are deduced from the classical pressure loss expressions [18-19] by Taylor's development. Therefore, the general form of Eqs. (2-3) can be given in the following way:

$$\begin{bmatrix} \dot{q}_d \\ \ddot{q}_d \\ \dot{q}_p \\ \ddot{q}_p \end{bmatrix} = \begin{bmatrix} 0 & 1 & 0 & 0 \\ -S_{dd} & -D_{dd} & -S_{dp} & -D_{dp} \\ 0 & 0 & 0 & 1 \\ -S_{pd} & -D_{pd} & -S_{pp} & -D_{pp} \end{bmatrix} \begin{bmatrix} q_d \\ \dot{q}_d \\ q_p \\ \dot{q}_p \end{bmatrix} + \begin{bmatrix} 0 \\ fNL_d(q_d, q_p) \\ 0 \\ fNL_p(q_d, q_p) \end{bmatrix} \qquad (7)$$

with the initial conditions: $q_{P|t=0} = 0$, $\dot{q}_{P|t=0} = 0$ and $q_{d|t=0} = 0$, $\dot{q}_{d|t=0} = 0$. Wherein, $S_{ij}$ and $D_{ij}$ are the stiffness and the damping coefficients between the $i$ and the $j$ DOF. $fNL_d(x,y)$ and $fNL_p(x,y)$ contains the nonlinear stiffness terms.

Hereafter, Eq. (7) will be written in state variable $\mathbf{x}^t = [\, q_d \,;\, \dot{q}_d \,;\, q_p \,;\, \dot{q}_p \,]$:

$$\dot{\mathbf{x}} = \mathbf{A}\,\mathbf{x} + \mathbf{F}_{NL2}(\mathbf{x} \otimes \mathbf{x}) + \mathbf{F}_{NL3}(\mathbf{x} \otimes \mathbf{x} \otimes \mathbf{x}) \qquad (8)$$

where $\mathbf{x}^t = [\, q_d \,;\, \dot{q}_d \,;\, q_p \,;\, \dot{q}_p \,]$, $\mathbf{A} = \begin{bmatrix} 0 & 1 & 0 & 0 \\ -S_{dd} & -D_{dd} & -S_{dp} & -D_{dp} \\ 0 & 0 & 0 & 1 \\ -S_{pd} & -D_{pd} & -S_{pp} & -D_{pp} \end{bmatrix}$.

$\otimes$ stands for the Kronecker product ( $[a, b] \otimes [c, d] = [a\,c,\, a\,d,\, b\,c,\, b\,d]$ ), $\mathbf{x} \otimes \mathbf{x}$ defines the basis of quadratic terms 2 and $\mathbf{x} \otimes \mathbf{x} \otimes \mathbf{x}$ defines the basis of cubic terms in the state space.

## 5. Solution procedure

The arising of steady oscillations in coupled nonlinear oscillators is classically associated to a supercritical Hopf bifurcation. At the fixed point also called the bifurcation point and denoted $\mathbf{x}_0$ hereafter a couple of complex conjugated eigenvalues associated to the system crosses the imaginary axes. Hence, this point becomes unstable and because of the nonlinearities of the system a stable limit cycle can occurs in its vicinity.

### 5.1. Unstable fixed point

The fixed point $\mathbf{x}_0$ is obtained by solving the non-linear static equations for a given set of parameter. It follows from Eq. (8) provided $\dot{\mathbf{x}} = \mathbf{0}$ that the only fixed point is $\mathbf{x}_0^t = [q_{d0};\, 0;\, q_{p0};\, 0] = [0;\, 0;\, 0;\, 0]$.

The stability is investigated by calculating the eigenvalues of the constant Jacobian matrix at the fixed point $\mathbf{x}_0$. It is possible to obtain the fourth-degree characteristic polynomial for which the roots are these eigenvalues, we set:

$$r^4 + \alpha r^3 + \beta r^2 + \gamma r + \delta = 0 \tag{9}$$

where:

$$\begin{aligned}
\alpha &= D_{pp} + D_{dd} \\
\beta &= D_{pp} D_{dd} - D_{pd} D_{dp} + S_{dd} + S_{pp} \\
\gamma &= - D_{dp} S_{pd} - D_{pd} S_{dp} + D_{dd} S_{pp} + D_{pp} S_{dd} \\
\delta &= S_{pp} S_{dd} - S_{pd} S_{dp}
\end{aligned} \tag{10}$$

Hence, depending on the operating parameters, three situations can occur:

If at least one of the real parts of the four roots would be greater than zero the system would be unstable in the vicinity of the fixed point.

If every real part would be less than zero the system is asymptotically stable. The motions of the membrane will tend to zero within a finite period of time.

If two roots have zero real part and the other two have negative real part, steady oscillations can occur.

The operation of the MMSE relies on the last case for which the roots of Eq. (9) at the fixed point can be expressed as two complex conjugates expressions:

$$\lambda_{Jc1} = -j\sqrt{\gamma/\alpha} = -j\,\omega_1\ ;\ \lambda_{Jc2} = j\sqrt{\gamma/\alpha} = +j\,\omega_1 \tag{11}$$

$$\lambda_{Js1} = \tfrac{1}{2}(-\alpha - j\sqrt{-\alpha^2 + 4\,\delta\alpha/\gamma}) = \alpha_2 - j\,\omega_2;\ \lambda_{Js2} = \tfrac{1}{2}(-\alpha + j\sqrt{-\alpha^2 + 4\,\delta\alpha/\gamma}) = \alpha_2 + j\,\omega_2 \tag{12}$$

where j is the complex number such as $j = \sqrt{-1}$.

Moreover, applying the Routh–Hurwitz criterion applied to the characteristic Eq. (9) results in the three following conditions for the stability:

C1: $\alpha > 0$ \hfill (13)

C2: $\alpha\beta - \gamma > 0$ \hfill (14)

C3: $\left(\dfrac{\gamma}{\alpha}\right)^2 - \beta\dfrac{\gamma}{\alpha} + \delta > 0$ \hfill (15)

These conditions are well known for the analyses of the FPSE [6]. De Monte et al. established the so called P4 theorem [4-5]. The evolutions of the terms of Eqs. (13-14) with respect to *Th/Tc* ratio and *p/p₀* ratio where $p_0$ is a reference pressure are plotted on Fig. 4.

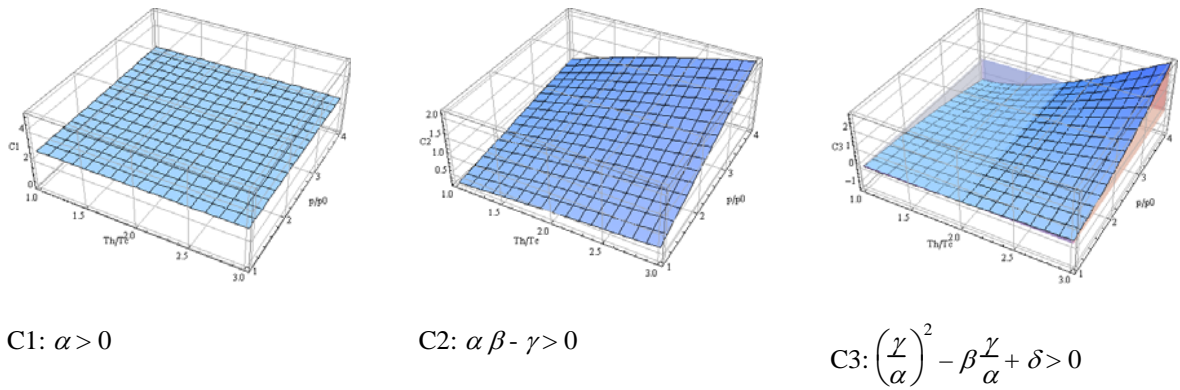

C1: $\alpha > 0$    C2: $\alpha\beta - \gamma > 0$    C3: $\left(\dfrac{\gamma}{\alpha}\right)^2 - \beta\dfrac{\gamma}{\alpha} + \delta > 0$

Fig. 4 Stability criteria for the behaviour of the engine.

As can be seen on the two graphics of the left, the first two conditions are usually met. The third condition defines an unstable region related to the highest temperature and mean pressure parameters. In the case of two pairs of complex conjugate eigenvalues, the evolution of their real parts is studied for increasing values of Th. The determining rule of the temperature for the apparition of the loss of stability is underlined in Fig. 5. As a result, a critical starting temperature can be defined.

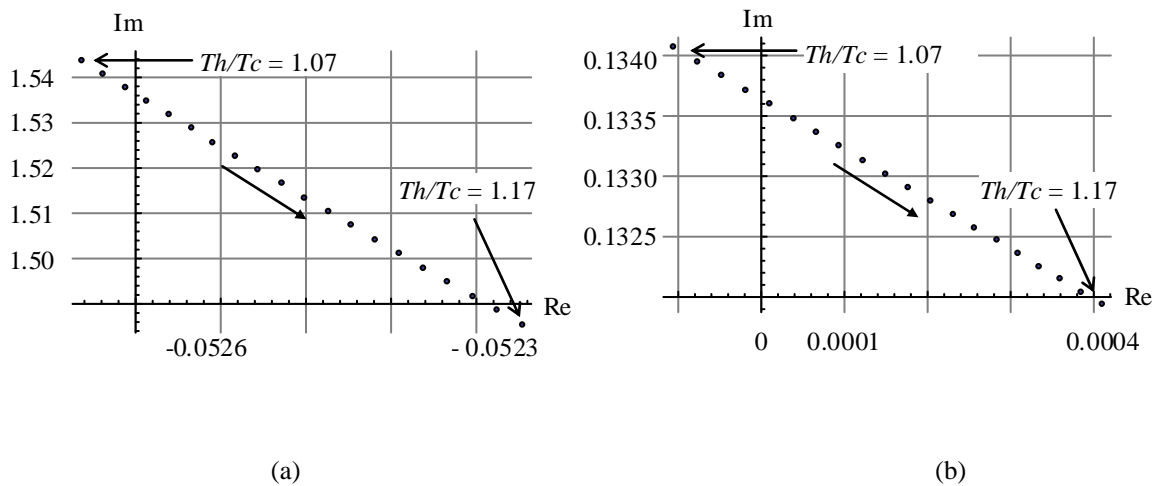

(a)    (b)

Fig. 5 Evolution of the system's poles with respect to temperature ratio *Th/Tc*, (a) first positive imaginary part pole and (b) first positive imaginary part pole.

The effect of the load of the system represented by the damping coefficient $\xi_p$ is reflected in Fig. 6 in which increasing $\xi_p$ leads to the stabilization of the system. Thus, the load may have to be controlled during the starting phase to make it easier.

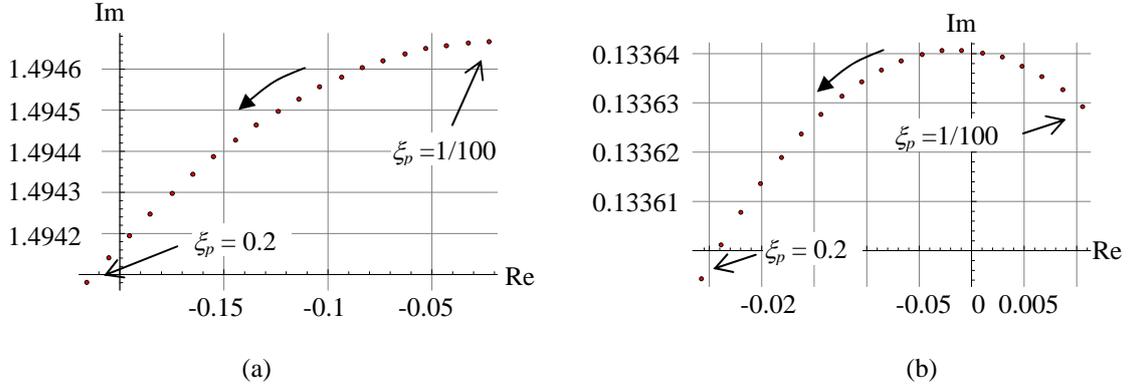

Fig. 6 Evolution of the system's poles with respect to $\xi_p$, (a) first positive imaginary part pole and (b) first positive imaginary part pole.

*5.2. Centre manifold approach*

The linear analysis can not represent the behaviour of the engine after the starting point. Therefore, another theory has to be used to study the post bifurcation behaviour. This section describes the application of the centre manifold [13] method to simplify the system defined by Eq. (8). Locally, the stability of the centre manifold is equivalent to the stability of the original system. Again, the bifurcation appears when one or several eigenvalues cross the imaginary axis in the complex plane with the variation of a control parameter (Fig. 5-6).

At the Hopf bifurcation point, one defines the coordinate transformation matrix $\mathbf{T}$ such as:

$$\begin{bmatrix} v_{c1} \\ v_{c2} \\ v_{s1} \\ v_{s2} \end{bmatrix} = \mathbf{T} \begin{bmatrix} q_d \\ \dot{q}_d \\ q_p \\ \dot{q}_p \end{bmatrix} \text{ and } \mathbf{T}^{-1} \mathbf{A} \mathbf{T} = \begin{bmatrix} \mathbf{J_c} & 0 \\ 0 & \mathbf{J_s} \end{bmatrix} = \begin{bmatrix} \alpha_1 & -\omega_1 & 0 & 0 \\ \omega_1 & \alpha_1 & 0 & 0 \\ 0 & 0 & \alpha_2 & -\omega_2 \\ 0 & 0 & \omega_2 & \alpha_2 \end{bmatrix} \quad (16)$$

where $\alpha_1$, $\omega_1$ and $\alpha_2$, $\omega_2$ are the real and imaginary parts of the complex eigenvalues defined in Eqs. (11-12). Note that at the bifurcation point $\alpha_1 = 0$. Thus, the previous system Eq. (8) can be written again in the form:

$$\begin{aligned} \dot{\mathbf{v}}_c &= \mathbf{J}_c \mathbf{v}_c + \mathbf{G}_2(\mathbf{v}_c, \mathbf{v}_s) + \mathbf{G}_3(\mathbf{v}_c, \mathbf{v}_s) \\ \dot{\mathbf{v}}_s &= \mathbf{J}_s \mathbf{v}_s + \mathbf{H}_2(\mathbf{v}_c, \mathbf{v}_s) + \mathbf{H}_3(\mathbf{v}_c, \mathbf{v}_s) \end{aligned} \quad (17)$$

where $\mathbf{J}_c$ et $\mathbf{J}_s$ have eigenvalues such as Re($\lambda_{Jc}$) = 0 and Re($\lambda_{Js}$) = $\alpha_2$ < 0. $\mathbf{G_2}$, $\mathbf{G_3}$, $\mathbf{H_2}$ and $\mathbf{H_3}$ are vector for which the two components are polynomials of degree 2 and 3 in the components of vectors $\mathbf{v}_c$ and $\mathbf{v}_s$ namely $v_{c1}$, $v_{c2}$ and $v_{s1}$, $v_{s2}$.

By considering the physically interesting case, it may be assumed that $\alpha_2$ is negative. $\mathbf{v}_s$ defines a stable subspace which is the linear approximation to the stable manifold. When we choose an initial condition on this stable manifold sufficiently close to the fixed point the solution curve will go toward the fixed point. In accordance to the parameters, an unstable subspace defined by the eigenvalues of $\mathbf{A}$ such as their real part is zero can exist. The center manifold theorem allows to represent locally the center manifold as { [$\mathbf{v}_c$, $\mathbf{v}_s$] such that $\mathbf{v}_s$ = $\mathbf{h}(\mathbf{v}_c)$, $\mathbf{h}(0) = 0$, $D\mathbf{h}(0) = 0$ } and consequently reduce the initial problem to a two dimensional one.

Substituting into the second line of Eq. (17) one obtains:

$$\dot{\mathbf{v}}_c = \mathbf{J}_c \mathbf{v}_c + \mathbf{G}_2(\mathbf{v}_c, \mathbf{v}_s) + \mathbf{G}_3(\mathbf{v}_c, \mathbf{v}_s) \tag{18}$$

$$D_{vc}(\mathbf{h}(\mathbf{v}_c))\dot{\mathbf{v}}_c = \mathbf{J}_s \mathbf{h}(\mathbf{v}_c) + \mathbf{H}_2(\mathbf{v}_c, \mathbf{h}(\mathbf{v}_c)) + \mathbf{H}_3(\mathbf{v}_c, \mathbf{h}(\mathbf{v}_c)) \tag{19}$$

Eq. (19) can not be solved analytically. However, it is possible to define an approximate solution of $\mathbf{h}$ by a power expansion without constant and linear terms. Thus, $\mathbf{h}$ is expressed as a power series in the components of $\mathbf{v}_c$ of degree $m$:

$$\mathbf{h}(\mathbf{v}_c) = \sum_{p\,=\,i+j=2}^{m} \sum_{j\,=\,0}^{p} \sum_{l\,=\,0}^{p} \mathbf{a}_{ij}\, v_{c1}^{i}\, v_{c2}^{j} \tag{20}$$

Replacing $\dot{\mathbf{v}}_c$ in Eq. (19) one obtains Eq. (21) form which the resolution provides the complex coefficients $\mathbf{a}_{ij}$ of the terms of $\mathbf{h}$.

$$D_{vc}(\mathbf{h}(\mathbf{v}_c))(\mathbf{J}_c \mathbf{v}_c + \mathbf{G}_2(\mathbf{v}_c, \mathbf{h}(\mathbf{v}_c)) + \mathbf{G}_3(\mathbf{v}_c, \mathbf{h}(\mathbf{v}_c))) = \mathbf{J}_s \mathbf{h}(\mathbf{v}_c) + \mathbf{H}_2(\mathbf{v}_c, \mathbf{h}(\mathbf{v}_c)) + \mathbf{H}_3(\mathbf{v}_c, \mathbf{h}(\mathbf{v}_c)) \tag{21}$$

As a result, the dynamic behaviour of the system is determined by a reduced two DOF system given by:

$$\dot{\mathbf{v}}_c = \mathbf{J}_c \mathbf{v}_c + \mathbf{G}_2(\mathbf{v}_c, \mathbf{h}(\mathbf{v}_c)) + \mathbf{G}_3(\mathbf{v}_c, \mathbf{h}(\mathbf{v}_c)) \tag{22}$$

*5.3. Normal form analysis*

The normal form analysis aims at a transformation of a system of nonlinear equations through a sequence of nonlinear-near identity transformations to eliminate as many nonlinear terms as possible. Those which cannot be removed are called the secular or the resonant terms. This simplest form of the equations is called the "normal form". The analysis of the dynamics of the normal forms yields a qualitative picture of the flows of each bifurcation type.

The centre manifold theorem has been applied to the initial system and henceforth we consider the resulting Eq. (22) detailed below with $\bar{v}_c$ the complex conjugate of $v_c$:

$$\begin{aligned} \dot{v}_c &= \alpha_1 v_c - \omega_1 \bar{v}_c + G_{2\text{-}1}(v_c, \bar{v}_c, \mathbf{h}(v_c, \bar{v}_c)) + G_{3\text{-}1}(v_c, \bar{v}_c, \mathbf{h}(v_c, \bar{v}_c)) \\ \dot{\bar{v}}_c &= \omega_1 v_c + \alpha_1 \bar{v}_c + G_{2\text{-}2}(v_c, \bar{v}_c, \mathbf{h}(v_c, \bar{v}_c)) + G_{3\text{-}2}(v_c, \bar{v}_c, \mathbf{h}(v_c, \bar{v}_c)) \end{aligned} \quad (23)$$

In which $G_{2\text{-}1}$ is the first component of the vector function $\mathbf{G_2}$, where, for $k = 0$ to 3 and $p = 0$ to 3:

$$G_{2\text{-}1}(v_c, \bar{v}_c, \mathbf{h}(v_c, \bar{v}_c)) + G_{3\text{-}1}(v_c, \bar{v}_c, \mathbf{h}(v_c, \bar{v}_c)) = \sum_{2<k+p<3} \alpha_{pk} v_c^k \bar{v}_c^p \quad (24)$$

$$G_{2\text{-}2}(v_c, \bar{v}_c, \mathbf{h}(v_c, \bar{v}_c)) + G_{3\text{-}2}(v_c, \bar{v}_c, \mathbf{h}(v_c, \bar{v}_c)) = \sum_{2<k+p<3} \beta_{pk} v_c^k \bar{v}_c^p \quad (25)$$

For the sake of clarity, Eq. (23) can be written as:

$$\begin{aligned} \dot{v}_c &= \alpha_1 v_c - \omega_1 \bar{v}_c + f(v_c, \bar{v}_c) \\ \dot{\bar{v}}_c &= \omega_1 v_c + \alpha_1 \bar{v}_c + g(v_c, \bar{v}_c) \end{aligned} \quad (26)$$

We observe that if we set $z = v_c - j\, \bar{v}_c$ and $\bar{z} = v_c + j\, \bar{v}_c$ the previous Eq. (26) is a pair of complex conjugate equations. Therefore, one single equation needs to be studied. A Hopf bifurcation is recognized here and the result of these successive transforms is well known [13].

$$\dot{v}_c = \alpha_1 v_c - \omega_1 \bar{v}_c + (a\, v_c - b\, \bar{v}_c)(v_c^2 + \bar{v}_c^2) \text{ and} \quad (27)$$

$$16\, a = (f_{xxx} + f_{xyy} + g_{xxy} + g_{yyy}) + \frac{1}{\omega_1}[f_{xy}(f_{xx} + f_{yy}) - g_{xy}(g_{xx} + g_{yy}) - f_{xx}g_{xx} + f_{yy}g_{yy}] \quad (28)$$

$$16\, b = (f_{xxy} + f_{yyy} - g_{xyy} - g_{xxx}) + \frac{1}{\omega_1}\, [2\, (f_{xx}^{\,2} + f_{xy}^{\,2}) + 5\, (f_{yy}^{\,2} + 5 g_{xx}^{\,2}) + 5 f_{xx}\, (f_{yy} - g_{xy})$$
$$+ 2 g_{xy}^{\,2} - f_{yy} g_{xy} + 5 g_{xx} g_{yy} + 2 g_{yy}^{\,2} - f_{xy}\, (g_{xx} + 5 g_{yy})]$$
(29)

In which $k_x$ is the partial derivative to the first order of a function $k$ with respect to the variable $x$ and $k_{xx}$ is the partial derivative to the second order of a function $k$ with respect to the variable $x$.

Finally, we have:

$$\dot{v}_c = \alpha_1\, v_c - \omega_1\, \bar{v}_c + (a\, v_c - b\, \bar{v}_c)\, (v_c^2 + \bar{v}_c^{\,2})$$
$$\dot{\bar{v}}_c = \omega_1\, v_c + \alpha_1\, \bar{v}_c + (b\, v_c + a\, \bar{v}_c)\, (v_c^2 + \bar{v}_c^{\,2})$$
(30)

We can transform Eq. (30) in polar form ($v_c = r \cos \theta$ and $\bar{v}_c = r \sin \theta$). Doing so, we obtain:

$$\dot{r} = \alpha_1\, r + a\, r^3$$
$$\dot{\theta} = \omega_1 + b\, r^2$$
(31)

The amplitude of the limit cycle can be assed by the analysis of the first of the previous equation. The non trivial solution establishes the existence condition for the limit cycle as $\frac{\alpha_1}{a} < 0$.

The stability of each fixed point can be studied by the sign of the Jacobian $J(r)$. From Eq. (31) $J(r) = \alpha_1 + 3\, a\, r^2$, therefore $J(r_{f1}) = \alpha_1$ and $J(r_{f2}) = -2\, \alpha_1$. Thus, if $\alpha_1$ is lower than zero and $a$ greater than zero, the only stable fixed point is $r_{f1} = 0$ and if $\alpha_1 > 0$ and $a < 0$, the unique stable fixed point is $r_{f2}$ and a limit cycle occurs.

These two last conditions added to the three previous ones ensure the existence of a stable limit cycle:

C4: $\alpha_1 > 0$ and C5: $a < 0$ (32)

Phase portraits of Fig. 7 shows the motion of the generalized coordinate for the expansion membrane ($q_d, \dot{q}_d$) in Fig. 7a and for the compression membrane ($q_p, \dot{q}_p$) in Fig. 7b. The results of a numerical resolution of Eq. 7, the resolution of Eq. (22) obtained from the centre manifold approach and the simplest expression of the result from the normal form analysis in Eq. (31) are shown in dotted (st), dashed-dotted (vc) and plain line (ap) respectively and present a good correlation for parameters close to the bifurcation point.

It is worth to notice that only the centre manifold approach followed by the normal form analysis allows an analytical expression of the solution.

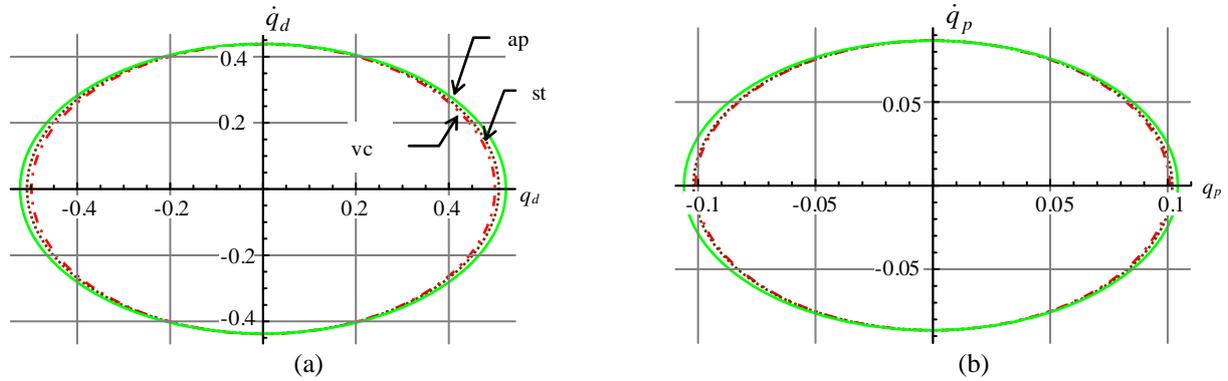

Fig. 7 Comparison of the results from straight resolution, centre manifold and normal form. (a) expansion membrane generalized coordinate position and velocity and (b) compression membrane generalized coordinate position and velocity.

## 6. Results and discussions:

In the previous section the results from the simplification method have been compared to the reference ones from a numerical resolution of Eq. (7). As a consequence, the approach is validated and is an efficient tool to study the effects of parameters on the motions of the membranes and eventually on the performances of the engine.

The classical control or variable parameters of a FPSE are: the heat source temperature; the heat sink temperature; the load and the internal pressure. The latter is barely controlled as it would require complex devices. In the specific case of a membrane Stirling engine when electric generation is considered, the mechanical load on the compression membrane is related to an electromechanical convertor device. Hence, we assume that it could be a part of the control parameters. The heat source temperature can be a control parameter for a quite large range of values whereas the heat sink temperature is related to external conditions, temperature specifically. Consequently in the following, we will focus on the effects of load and heat source variations.

For each simulation, the Hopf bifurcation point and the value of the critical ratio of the hot to cold temperature $Th/Tc$ are computed. The study is provided near the bifurcation point by choosing the parameter value $Th$ not greater than 1% of its critical value. It is obvious that a lower value than the critical one does not match the condition of Eqs. (13-15).

Table 2 Evolution of the limit cycles near the bifurcation point with respect to $\xi_p$.

| $\xi p$ | 0.04 | 0.06 | 0.08 | 0.1 | 0.12 | 0.14 |
|---|---|---|---|---|---|---|
| $Tcriti/Tc$ | 1.09748 | 1.67958 | 2.08963 | 2.41272 | 2.68125 | 2.91167 |
| *Thtest=1.005\*Thcrit* | | | | | | |
| Limit cycle radius | 0.83409 | 1.17243 | 1.41678 | 1.60137 | 1.74944 | 1.87276 |
| *Thtest=1.01\*Thcrit* | | | | | | |
| Limit cycle radius | 1.18367 | 1.67068 | 2.02273 | 2.28975 | 2.50421 | 2.68354 |
| *Thtest=1.03\*Thcrit* | | | | | | |
| Limit cycle radius | 2.07986 | 2.98424 | 3.64447 | 4.15197 | 4.56447 | 4.91321 |

It is observed that the level amplitude is a complex problem. The evolution of limit cycle amplitude is not linear with the evolution of a specific parameter. It turns out that the increasing of the load leads to a reduction of the amplitude of the compression membrane whereas the expansion membrane amplitude increases. This is further reflected in Figs. 8–10.

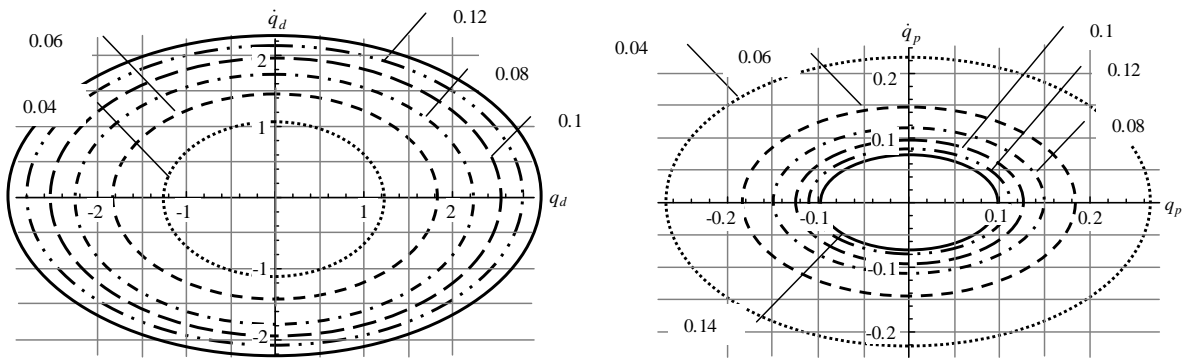

Fig. 8 Phase portraits for $Th = 1.005\ Tc$ and various dissipative coefficient $\xi_p$.

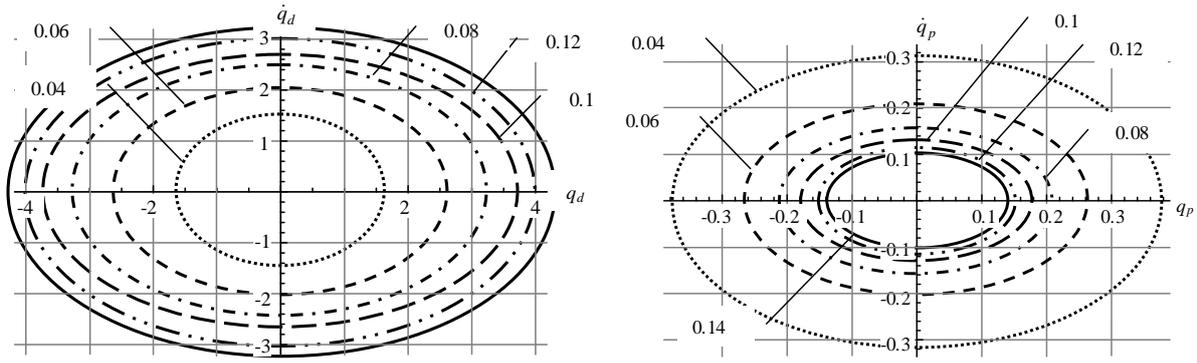

Fig. 9 Phase portraits for $Th = 1.01\ Tc$ and various dissipative coefficient $\xi_p$.

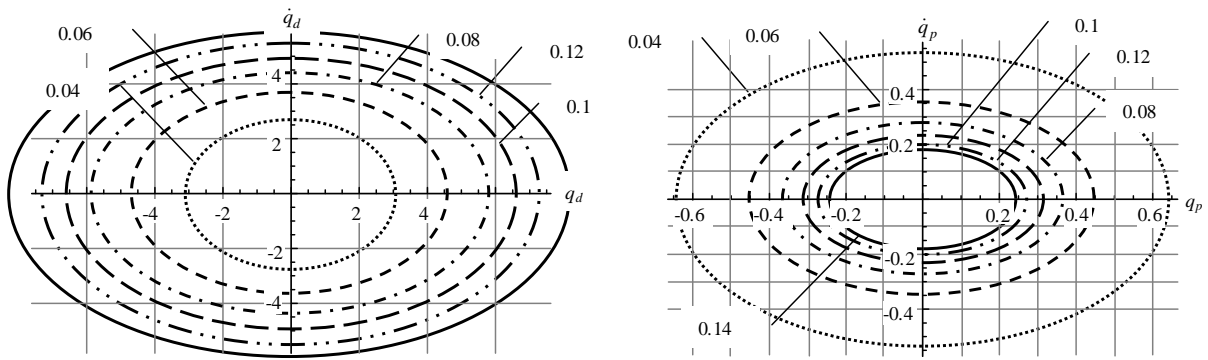

Fig. 10 Phase portraits for $Th = 1.03\ Tc$ and various dissipative coefficient $\xi_p$.

The evaluations of the efficiency and the output power of the system are given in table 3 below. As the upper temperature $Th$ is varying, the relative efficiency defined as the ratio of the efficiency with the Carnot efficiency $\eta_C = 1\text{-}Tc/Th$, allows a fair comparison between the different cases studied.

It is worth of noting that the ability of the engine to operate as a prime mover relies on the phase between the compression and the expansion membrane motions. If the damping coefficient $\xi_p$ is smaller than 0.08, the compression membrane doesn't lag behind the expansion any more thus leading to a reverse behaviour of the Stirling cycle which is a cooler or heat pump. The sign of the mean output power changes in this case.

Table 3 Evolution of the performances with respect to $Th$ and $\xi_p$.

| $\xi p$ | 0.04 | 0.06 | 0.08 | 0.1 | 0.12 | 0.14 |
|---|---|---|---|---|---|---|
| $Tcriti/Tref$ | 1.10 | 1.68 | 2.09 | 2.41 | 2.68 | 2.91 |

*Thtest*=1.005**Thcrit*

| Frequency (Hz) | *225* | *259* | 287 | 309 | 327 | 342 |
|---|---|---|---|---|---|---|
| Efficiency (%) | --- | --- | 29.25 | 30.79 | 31.68 | 32.25 |
| Relative efficiency | --- | --- | 55.85 | 52.41 | 50.37 | 48.98 |
| Output power (mW) | *-27* | *-36* | 2 | 55 | 113 | 171 |

*Thtest*=1.01**Thcrit*

| Frequency (Hz) | *250* | *303* | 345 | 378 | 405 | 427 |
|---|---|---|---|---|---|---|
| Efficiency (%) | --- | --- | 29.32 | 30.84 | 31.71 | 32.28 |
| Relative efficiency | --- | --- | 55.71 | 52.30 | 50.28 | 48.91 |
| Output power (mW) | *-61* | *-80* | 17 | 155 | 310 | 470 |

*Thtest*=1.03**Thcrit*

| Frequency (Hz) | *314* | *421* | 502 | 566 | 618 | 662 |
|---|---|---|---|---|---|---|
| Efficiency (%) | --- | --- | 29.55 | 31.02 | 31.86 | 32.40 |
| Relative efficiency | --- | --- | 55.19 | 51.9 | 49.94 | 48.61 |
| Output power (mW) | *-237* | *-253* | 303 | 1115 | 2056 | 3063 |

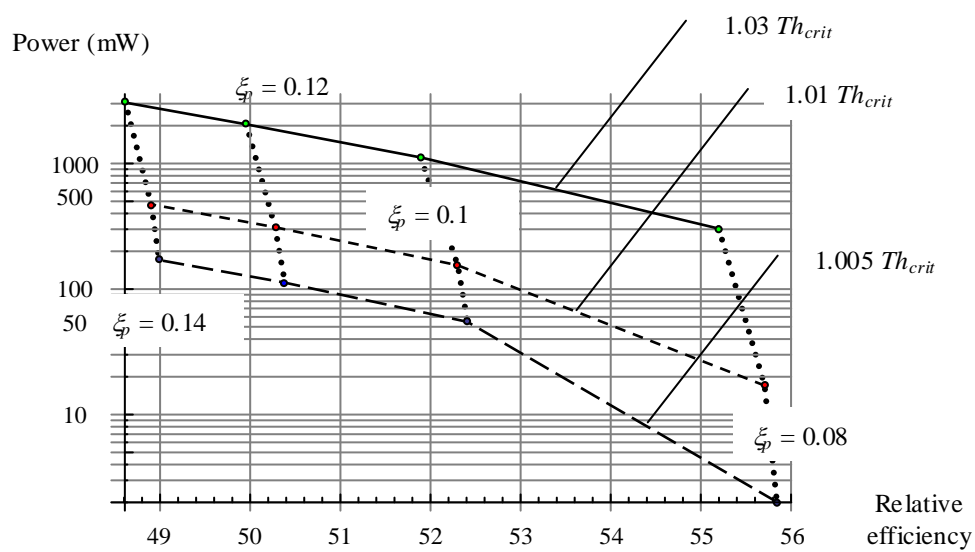

Fig. 11 Evolution of the relative efficiency - power for various upper temperature and dissipative coefficient.

From Fig. 11, it can be seen that the efficiency of the engine is strongly dependant on the load. From the small amplitude of the compression membrane a smaller compression ratio can be inferred but higher operating frequencies involve higher output power. However, larger amplitude for the expansion membrane associated to higher frequencies means higher flow rate through the heat exchangers and regenerator. Hence, the pressure losses in addition to viscous dissipation at the expansion membrane reduce the efficiency.

**7. Conclusion**

A non-linear model for the analysis of the operation of a microminiature Stirling engine has been developed. For the understanding and the design of the engine, the nonlinear behaviour of the membrane as well as the thermodynamic of the engine has been considered. The existence of a Hopf bifurcation conditions the operation and performances. Moreover, this paper presents the centre manifold– normal form approach in order to obtain analytical conditions and expression for the limit cycle amplitude. This approach simplifies the analysis and the design by reducing the order of the dynamical system, while retaining the essential features of the dynamic behaviour near the Hop bifurcation point. Tough it is based on a highly idealized thermodynamic cycle the semi analytical model is representative of the global behaviour of the engine and thus provides an efficient tool for preliminary design in comparison with extensive parametric design studies. It outlines the effect of control parameters on swept volume through the amplitude of the membrane flexions. Therefore, ideal performances can be evaluated and optimized.

**References**

[1] W. Beale "Stirling cycle type thermal device." US Patent 3552120, 1971.

[2] N.C. Chen, F.P. Griffin, "A review of Stirling engine Mathematical models", Oak Ridge National Laboratory for the US Department of Energy, 1983.

[3] Schmidt G. Theorie der Lehmannschen calorischen maschine Zeit Des Vereines deutsch Ing, 1871.

[4] F. De Monte, G. Benvenuto, "Reflections on Free-Piston Stirling Engines, Part 1: Cyclic Steady Operation", Journal of Propulsion and Power, vol.14, p. 499-508, 1998.

[5] F. De Monte, G. Benvenuto, "Reflections on Free-Piston Stirling Engines, Part 2: Stable Operation", Journal of Propulsion and Power, vol.14, p. 509-518, 1998.

[6] D.M. Berchowitz, "Operational Characteristics of Free-piston Stirling Engines", Proceedings of the 23rd Intersociety Energy Conversion Engineering, 1988.

[7] L. Bowman, D.M. Berchowitz and al., "Microminiature stirling cycle cryocoolers and engines", US Patent 05749226, 1994

[8] J. McEntee, L. Bowman, "Oscillating Diaphragms", Proceedings of the International Conference on Modeling and Simulation of Microsystems 2, 1999.

[9] M. Haterboucha, R. Benamar, "The effects of large vibration amplitudes on the axisymmetric mode shapes and natural frequencies of clamped thin isotropic circular plates. Part I: iterative and explicit analytical solution for non-linear transverse vibrations", Journal of Sound and Vibration 265, p. 123–154, 2003

[10] Rogdakis E.D., Bormpilas N.A. and al., "A thermodynamic study for the optimization of stable operation of free piston Stirling engines", Energy Conversion and Management, vol 45, p. 575-593, issue 4, 2004.

[11] A.H. Nayfeh, D.T. Mook, "Nonlinear oscillations", Wiley, New York, 1979

[12] J.E. Marsden, M. McCracken, "The Hopf Bifurcation and its Applications", Applied Mathematical Sciences. Vol. 19, Springer, Berlin, 1976

[13] J. M. Guckenheimer and P. Holmes, "Nonlinear Oscillation, Dynamical Systems and Bifurcation of Vector Fields", New York: Springer-Verlag, 1983

[14] M. Haterboucha, R. Benamar, "The effects of large vibration amplitudes on the axisymmetric mode shapes and natural frequencies of clamped thin isotropic circular plates. Part II: iterative and explicit analytical solution for non-linear coupled transverse and in-plane vibrations", Journal of Sound and Vibration 277, p. 1–30, 2004

[15] M.L. Kadiri and R. Benamar, "Improvement of the semi-analytical method for determining the geometrically non-linear response of thin straight structures. Part I: Application to clamped-clamped and simply supported-clamped beams", Journal of Sound and Vibration 249(2), p. 263–305, 2002

[16] M.L. Kadiri and R. Benamar, "Improvement of the semi-analytical method for determining the geometrically non-linear response of thin straight structures. Part II: First and second non-linear mode shapes of fully clamped rectangular plates", Journal of Sound and Vibration 257(1), p. 19–62, 2002

[17] M.L. Kadiri and R. Benamar, "Improvement of the semi-analytical method for determining the geometrically non-linear response of thin straight structures. Part III: Steady state periodic forced response of rectangular plates", Journal of Sound and Vibration 264(1), p. 1–35, 2003

[18] Kays, W.M., and London, A.L., "Compact Heat Exchangers," McGraw-Hill, 1964

[19] Martini WR. "Stirling engine design manual", report no. NASA CR-135382. In: M.J. Collie, editor.

**Appendix**

*A1. Semi analytical method for the compression membrane*

The strategy to obtain the analytical models of the membranes is introduced with the simple example of the compression membrane. The geometry of the compression membrane is described on Fig. 12.

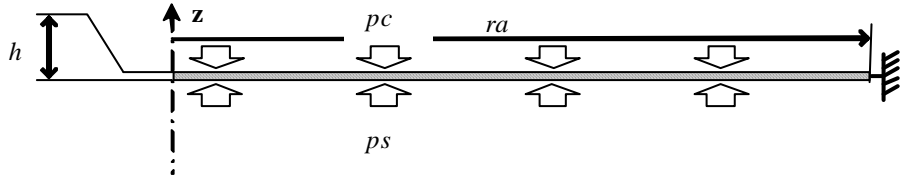

Fig. 12 Definition of the compression membrane parameters.

The equilibrium equations are obtained using the Hamilton's principle:

$$\delta \int_{t_1}^{t_2} (E_c - E_d - W_{ext})\, dt = 0 \quad (A1)$$

where, $E_c$ is the kinetic energy, $E_d$ the strain energy and $W_{ext}$ is the work of the external forces related to the pressure in the considered case. $t_1$ and $t_2$ are times such as no deflexion of the membrane occurs.

Denoting $w^*(r^*)$ for the out of plane midplane displacement of the membrane located at the radius $r^*$, $E_d$ and $W_{ext}$ can be expressed as follow:

$$E_c = \pi \rho h \int_0^{ra} \dot{w}^{*2} r^* dr^* \quad (A2)$$

$$E_d = \pi D \int_0^{ra} \left[ w^*_{r^*r^*}{}^2 + \frac{1}{r^{*2}} w^*_{r^*}{}^2 + \frac{3}{h^2} w^*_{r^*}{}^4 \right] r^* dr^* \quad (A3)$$

where, $D = \frac{E h^3}{12(1-\nu^2)}$ is the bending stiffness of the plate, and $E$, $\nu$ and $\rho$ are the Young's modulus, the Poisson ratio and the density of the plate material, whereas $h$ the thickness of the membrane. $w^*_{r*}$ is the partial derivative to the first order of a function $w^*$ with respect to the variable $r^*$ and $w^*_{r*r*}$ is the partial derivative to the second order of a function $w^*$ with respect to the variable $r^*$.

The total strain energy, $E_d$ of the circular plate is given as the sum of the strain energy due to bending and the membrane strain energy induced by large deflections. In the case of axisymmetric vibrations, the bending strain energy is given by:

$$E_d = E_l + E_{nl} = \pi D \int_0^{ra} \left[ w^{*^2}_{r*r*} + \frac{1}{r^{*2}} w^{*^2}_{r*} \right] r^* dr^* + \frac{3\pi D}{h^2} \int_0^{ra} w^{*^4}_{r*} r^* dr^* \qquad (A4)$$

In the considered application, the work of the external pressure forces is:

$$W_{ext} = 2\pi p \int_0^{ra} w^* r^* dr^* \qquad (A5)$$

where $p$ stands for an uniform pressure applied on the surfaces of the membrane. In the case of the compression membrane and with respect to the notation of Fig. 1, $p = ps\text{-}pc$.

If the space and time functions are supposed to be separable, the transverse displacement can be expanded in the form of finite series of n basis functions $W^*_i(r)$. Hence, with the usual summation convention for repeated indices is used over the range $[1, n]$, we set $w^*(r,t) = q_i(t) W^*_i(r)$

The discretization of the total strain, kinetic energy and force work expressions is made by substituting the expression for $w^*(r,t)$ in Eqs. (A2-A5). This leads to the following expressions:

$$\begin{aligned} E_l &= \tfrac{1}{2} q_i q_j k^*_{ij} \\ E_{nl} &= \tfrac{1}{2} q_i q_j q_k q_l b^*_{ijkl} \\ E_c &= \tfrac{1}{2} \dot{q}_i \dot{q}_j m^*_{ij} \\ W_{ext} &= p\, q_i f^*_i \end{aligned} \qquad (A6)$$

where $k^*_{ij}$, $b^*_{ijkl}$ and $m^*_{ij}$ are the stiffness tensor, the non-linearity stiffness tensor, the mass tensor and the force tensor respectively, given by:

$$k^*_{ij} = 2\pi D \int_0^{ra} \left[ \frac{d^2W^*_i}{dr^{*2}} \frac{d^2W^*_j}{dr^{*2}} + \frac{1}{r^{*2}} \frac{dW^*_i}{dr^*} \frac{dW^*_j}{dr^*} \right] r^* dr^*$$

$$b^*_{ijkl} = 2\pi \frac{3D}{h^2} \int_0^{ra} \left[ \frac{dW^*_i}{dr^*} \frac{dW^*_j}{dr^*} \frac{dW^*_k}{dr^*} \frac{dW^*_l}{dr^*} \right] r^* dr^*$$

$$m^*_{ij} = 2\pi \rho h \int_0^{ra} W^*_i W^*_j r^* dr^*$$

$$f^*_i = 2\pi \int_0^{ra} W^*_i r^* dr^*$$

In which $W^*(r) = A\, J_0(\beta r^*) + B\, Y_0(\beta r^*) + C\, I_0(\beta r^*) + D\, K_0(\beta r^*)$ where $J_0(x)$, $I_0(x)$ et $Y_0(x)$, $K_0(x)$ stands for the Bessel functions and modified Bessel function respectively of the first and the second kind and $\beta^4 = \frac{\omega^2 rh}{D}$.

Replacing $E_c$, $E_d$ and $W_{ext}$ by their discretized expressions in the energy condition Eq. (A1), calculating the derivatives with respect to the $q_i$'s, and taking into account the properties of symmetry of the tensors leads to the following:

$$\ddot{q}_i m^*_{in} + q_i k^*_{in} + 2\, q_i q_j q_k\, b^*_{ijkn} = f^*_n \tag{A7}$$

Eqs. (A7) represent a set of n non-linear algebraic equations relating the $n$ time functions $q_i$.

For a single vector basis we obtain:

$$\boxed{\ddot{q}_1 m^*_{11} + q_1 k^*_{11} + 2\, q_1^3\, b^*_{1111} = p f^*_1} \tag{A8}$$

where $\ddot{q}_1$ and $q_1$ are the acceleration and displacement response of the DOF, respectively

Consequently, Eq. (A8) can be seen as a conservative Duffing's equation.

*A2. Semi analytical method for the expansion membrane*

The expansion membrane is a different case because of its architecture. It is composed of two circular plates linked by a central rigid circular stem of radius $b$. The plates have two different radius $a_1$ and $a_2$ respectively (Fig. 13).

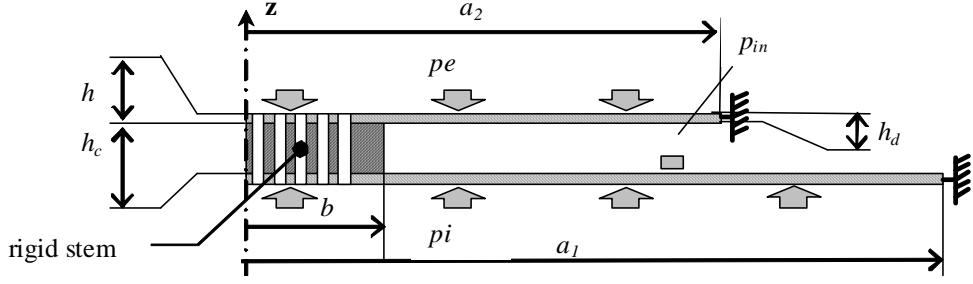

Fig. 13 Definition of the expansion membrane parameters

The kinetic and the total strain energy of the expansion membrane take into account the geometry. In order to simplify the presentation, the material and thicknesses of the two circular plates are supposed to be the same. Nevertheless, the methodology can be applied to a more general case.

The transverse displacement of the lower plate which radius is $a_1$ is denoted $w1^*(r,t)$ whereas the transverse displacement of the upper plate is $w2^*(r,t)$.

In the case of axisymmetric vibrations, the kinetic and the bending strain energy are given by Eqs. (A9) and (A10) below:

$$E_d = \pi D \int_b^{a_1} \left[ {w1^*_{r^*r^*}}^2 + \frac{1}{r^{*2}} {w1^*_{r^*}}^2 + \frac{3}{h^2} {w1^*_{r^*}}^4 \right] r^* dr^* + \pi D \int_b^{a_2} \left[ {w2^*_{r^*r^*}}^2 + \frac{1}{r^{*2}} {w2^*_{r^*}}^2 + \frac{3}{h^2} {w2^*_{r^*}}^4 \right] r^* dr^* \quad (A9)$$

$$E_c = \pi \rho h \int_b^{a_1} {\dot{w1}^*}^2 r^* dr^* + \pi \rho h \int_b^{a_2} {\dot{w2}^*}^2 r^* dr^* + \frac{1}{4} m_c \left( {\dot{w1}^*}\big|_b^2 + {\dot{w2}^*}\big|_b^2 \right) \quad (A10)$$

The mechanical works done by the pressure forces are:

$$W1_{ext} = -2\pi\, pHC \left[ \int_b^{a_1} w1^* r^* dr^* + \frac{1}{2} b^2 w1^*\big|_{b^-} \right] = -pHC\, f1_d^* \quad (A11)$$

$$W2_{ext} = 2\pi\, pIC \left[ \int_b^{a_2} w2^* r^* dr^* + \frac{1}{2} b^2 w2^*\big|_{b^-} \right] = pIC\, f2_d^* \quad (A12)$$

$$W3_{ext} = 2\pi\, p_{atm} \int_b^{a_1} w1^* r^* dr^* = p_{atm}\, f3_d^* \quad (A13)$$

$$W4_{ext} = -2\pi\, p_{atm} \int_b^{a_2} w2^* r^* dr^* = -p_{atm}\, f4_d^* \quad (A14)$$

Using the previous approach, the transverse displacements can be expanded in the form of finite series of $n$ basis functions $W^*_i(r)$. Hence, with the usual summation convention for repeated indices is used over the range $[1, n]$, we set $w^*(r,t) = q_i(t)\, W^*_i(r)$.

In the case of the expansion membrane, the local equilibrium equation for each of the circular plates is the same as for the compression membrane, though the boundary conditions are different.

For the external edges we set:

$$W1^*_{r*}(a_1) = 0,\ W1^*(a_1) = 0,\ W2^*_{r*}(a_2) = 0,\ W2^*(a_2) = 0 \tag{A15}$$

Two other types of condition are related to the flexural moment and the zero slope at the central stem $Wz^*_{r*}$ with $z = 1, 2$.

$$2\pi\, b\, D \left[ W1^*_{r*r*r*} + \frac{1}{r^*} W1^*_{r*r*} - \frac{1}{r^{*2}} W1^*_{r*} \right]_{\Big|_b} + 2\pi\, b\, D \left[ W2^*_{r*r*r*} + \frac{1}{r^*} W2^*_{r*r*} - \frac{1}{r^{*2}} W2^*_{r*} \right]_{\Big|_b} = mc\, \ddot{W1}^*_{\Big|} \tag{A16}$$

$$W1^*_{r*}(b) = 0 \tag{A16a}$$

$$W2^*_{r*}(b) = 0 \tag{A16b}$$

$$W1^*(b) = W2^*(b) \tag{A16c}$$

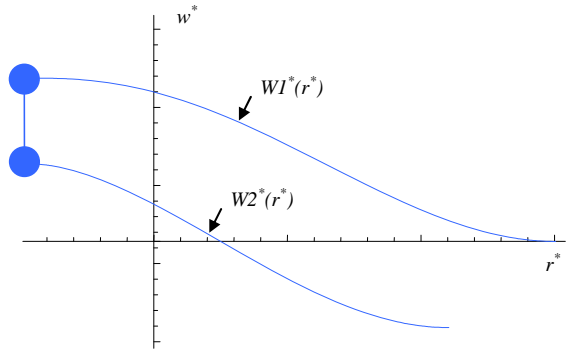

Fig. 14 Scheme of the first linear flexion mode of the expansion membrane.

Fig. 14 shows a scheme of the first linear flexion mode for the expansion membrane. The central rigid stem is represented by two punctual masses.

In the following table the results form a finite element analysis and the analytical approach for $\frac{a1}{b}$ ratios varying from 5 to 100 demonstrate a perfect correlation which allow the validation of the analytical approach for the evaluation of the spectral basis.

Table A1

| Frequency of mode 1 (Hz) | 100 | 40 | 20 | 10 | 5 |
|---|---|---|---|---|---|
| Finite Element Analysis | 155.401 | 154.724 | 149.128 | 117.543 | 61.196 |
| Analytical approach | 155.397 | 154.666 | 148.886 | 117.051 | 60.986 |
| Discrepancy (%) | 0.002 | 0.037 | 0.162 | 0.418 | 0.343 |

Using the previously described approach for the discretized expressions of the energy terms in the Hamilton's principle, the set of non-linear algebraic equations for the expansion membrane can be obtained in the same way.